\documentclass[aps,11pt]{article}
\begin{document}

\title{Lagrangian description of the radiation damping}

\author{P.M.V.B. Barone and A.C.R. Mendes
\thanks{\noindent e-mail:barone@fisica.ufjf.br, albert@fisica.ufjf.br}\\
Departamento de F\'{\i}sica, Universidade Federal de
Juiz de Fora, \\36036-330, Juiz de Fora, MG, Brasil}

\maketitle
\begin{abstract}
We present a Lagrangian formalism to the dissipative system of a charge interacting with its own radiation field, which gives rise to the radiation damping \cite{Heitler}, by the indirect representation doubling the phase-space dimensions. 
\end{abstract}

\maketitle

\section{Introduction}

\label{intro}
The study of dissipative systems in quantum theory is of strong interest and relevance either for fundamental reasons \cite{Zurek} and for its practical applications \cite{Amir1,NMR}. The explicit time dependence of the Lagrangian and Hamiltonian operators introduces a major difficulty to this study since the canonical commutation relations are not preserved by time evolution. Then different approaches have been used in order to apply the canonical quantization scheme to dissipative systems (see, for instance, \cite{Flavio,Barone}).

One of these approaches is to focus on an isolated system composed by the original dissipative system plus a reservoir. One start from the beginning with a Hamiltonian which describes the system, the bath and the system-bath interaction. Subsequently, one eliminates the bath variables which give rise to both damping and fluctuations, thus obtaining the reduced density matrix \cite{FeynmanVernon,Amir1,Amir2,Amir3,Barone}.

Another way to handle the problem of quantum dissipative systems is to double the phase-space dimensions, so as to deal with an effective isolated system composed by the original system plus its time-reversed copy (indirect representation) \cite{Banerjee,Feshbach}. The new degrees of freedom thus introduced may be represented by a single equivalent (collective) degree of freedom for the bath, which absorbs the energy dissipated by the system.

The study of the quantum dynamics of an accelerated charge is appropriated to use the indirect representation since it loses the energy, the linear momentum, and the angular momentum carried by the radiation field \cite{Heitler}. The effect of these losses to the motion of charge is know as radiation damping \cite{Heitler}.

The reaction of a classical point charge to its own radiation was first discussed by Lorentz and Abraham more than one hundred years ago, and never stopped being a source of controversy and fascination \cite{Becker,Lorentz}. Nowadays, it is probably fair to say that the most disputable aspects of the Abraham-Lorentz theory, such as self-acceleration and preacceleration, have been adequately understood. Self-acceleration refers to classical solutions where the charge is under acceleration even in the absence of an external field. Preacceleration means that the charge begins to accelerate before the force is actually applied.

The process of radiation damping is important in many areas of electron accelerator operation \cite{Walker}, as in recent experiments with intense-laser relativistic-electron scattering at laser frequencies and field strengths where radiation reaction forces begin to become significant \cite{Hartemann,Bula}.

The purpose of this letter is to present a Lagrangian formalism to the study of quantum dynamics of accelerated charge, yielding an {\it effective} isolated system, where the canonical commutation relations are preserved by time evolution. In Section \ref{sec:1} we briefly review the equation of motion of the radiation damping and aspects of the solutions to the equation of motion. In section \ref{sec:2} we present a Lagrangian description to the radiation damping by the indirect representation, doubling the phase-space dimensions. Section \ref{sec:3} contains the concluding remarks. 

\section{The equation of motion}
\label{sec:1}
The derivation of an exact expression for the radiation damping force has long been an outstanting problem of classical electrodynamics \cite{Flavio,Heitler,Becker,Lorentz,Hartemann,Rohrlich1,Landau,Rohrlich2}. In the classic derivation given by Lorentz and Abraham \cite{Becker,Lorentz}, which relies on energy-momentum conservation, the self-elec\-tro\-mag\-ne\-tic-energy and momentum of a charged rigid sphere are derived for an accelerated motion. In this first-order ap\-pro\-xi\-ma\-ti\-on, this de\-ri\-va\-ti\-on yields the well-known Abraham -Lorentz force which depends on the second time de\-ri\-va\-ti\-ve of the particle velocity of mass $m$ and charge $e$:
\begin{equation}\label{01}
m\left( \frac{d \vec v}{dt} -\tau_0 \frac{d^2 \vec v}{dt^2} \right) =\vec F,
\end{equation}
where $\tau_0=2e^2/3mc^3$, $c$ is the velocity of light, $\vec v=d\vec r/dt$ denotes the velocity of the charge, and $\vec F$ is the external force. 

A fully relativistic formulation of the equation of motion was only achieved in 1938 by Dirac in his classic paper \cite{Dirac}, where the Lorentz-Dirac equation reads
\begin{equation}\label{02}
ma^{\mu}=\frac{e}{c}F^{\mu\nu}u_{\nu} +\Gamma^{\mu},
\end{equation}
with
\begin{equation}\label{03}
\Gamma^\mu \equiv \frac{2e^2}{3c^3}\left(\dot a^\mu-a^\lambda a_\lambda \frac{u^\mu}{c^2}\right),
\end{equation}
where the charge world line $z_\mu (\tau)$ is parametrized by its proper time $\tau$, and $u_\mu = dz/d\tau$, $a_\mu =du_\mu /d\tau$, and $\dot a_\mu =da_\mu /d\tau$. Greek indices range from 0 to 3, and the diagonal metric of Minkowski space is $(-1,1,1,1).$ The term $(e/c)F^{\mu\nu}u_\nu$ in Eq.(\ref{02}) is the Lorentz force due to the external field $F^{\mu\nu}$. In addition, $\Gamma^\mu$ represents the effect of radiation \cite{Rohrlich1}.

The equation (\ref{01}) can be critized on the grounds that it is second order in time, rather than first, and therefore runs counter to the well-known requirements for a dynamical equation of motion. This difficulty manifests itself immediately in runaway (self-accelerated) solutions. If the external force is zero, with the help of the integrating factor $e^{t/\tau_0}$, it is obvious that Eq.(\ref{01}) has two possible solutions,
\begin{equation}\label{04}
\dot{\vec v}(t)=\left\{ \begin{array}{l}
0 \\
\dot{\vec v}(0)e^{t/\tau_0}.
\end{array}
\right.
\end{equation}
Only the first solution is rasonable. 

However, there are a particular choice for $\dot{\vec v}(0)$ where the second solution in Eq.(\ref{04}) disappear, that is the Dirac's asymptotic condition on the vanishing of the acceleration for an asymptotically free particle \cite{Dirac}. In this case, the solution of Eq.(\ref{01}), with the help of the integrating factor $e^{t/\tau_0}$, for a rather general time-dependent force $\vec F(t)$ reads
\begin{equation}\label{05}
m\frac{d\vec v}{dt}=\int_0^\infty e^{-s}\vec F(t+\tau_0 s)ds.
\end{equation}
In fact, if $\vec F(t)$ vanishes identically for a large value of $t$, then Eq.(\ref{05}) shows that the acceleration also vanishes for a large value of $t$ and therefore solution (\ref{05}) is not self-accelerating. But, unfortunately, and althogh mathematically correct, this approach leads to preacceleration. The violation of causality implied by preacceleration is particularly disappointing since the Lorentz-Dirac equation (\ref{02}) can be derived by using only retarded fields \cite{Villaroel}. The existence of preacceleration is not a consequence of the presence  the time derivative of the acceleration in (\ref{01}), but of the method through which the solution has been obtained \cite{Villaroel1}.

\section{Indirect Lagrangian representation of the radiation damping}
\label{sec:2}

The inverse problem of variational calculus is to construct the Lagrangian from the equations of motion. Different Lagrangian representations are obtained from the direct and indirect approaches \cite{Santilli}. In the direct representation as many variables are introduced as there are in the equations of motion. The equation of motion corresponding to a coordinate $q$ is related with the variational derivative of the action with respect to the same coordinate. Whereas, in the indirect representation, the equation of motion is suplemented by its time-reversed image. The equation of motion with respect to the original variable then corres\-ponds to the variational derivative of the action with res\-pect to the image coordinate and vice versa \cite{Feshbach,Bateman}.

In the indirect approach we consider equation (\ref{01}) along with its time-reversed copy
\begin{equation}
\label{06}
m\left( \frac{d \vec {\bar v}}{dt} +\tau_0 \frac{d^2 \vec {\bar v}}{dt^2} \right) =\vec{\bar  F},
\end{equation}
where $\vec {\bar v}=d\vec{\bar r}/dt$ is the velocity of the image system, which appears in fact to be the {\it time reversed} $(\tau_0 \rightarrow -\tau_0 )$ of (\ref{01}).

Thus the variation of the action $S$ for equations of motion (\ref{01}) and (\ref{06}), in term of the coordinates, must then be
\begin{eqnarray}\label{07}
\delta S =\int_{t_1}^{t_2}dt && \left[ m\left(\frac{d}{dt}\dot{\vec r}-\tau_0 \stackrel{\ldots}{\vec r}+\frac{\partial V}{\partial\vec{\bar r}}\right). \delta\vec {\bar r}\right. \nonumber\\ 
&+& \left. m\left(\frac{d}{dt}\dot{\vec {\bar r}}+\tau_0 \stackrel{\ldots}{\vec {\bar r}}+\frac{\partial V}{\partial\vec r}\right). \delta\vec r \right],
\end{eqnarray}
where $V\equiv V(\vec r, \vec {\bar r})$ is the potential energy  with $\frac{\partial V}{\partial\vec r}=-\vec {\bar F}$ and $\frac{\partial V}{\partial\vec{\bar r}}=-\vec F$.
>From (\ref{07}), equation (\ref{01}) is obtained by varying $S$ with $\vec {\bar r}$ whereas (\ref{06}) follows from varying $S$ with $\vec r$. Since the equations of motion for $\vec r$ and $\vec {\bar r}$ follow as Euler-Lagrangian equations of motion for $\vec {\bar r}$ and $\vec r$ respectively, the method is called the indirect method. By descarding the surface terms, we get from (\ref{07}):
\begin{equation}
\label{08}
\delta S =-\delta \int_{t_1}^{t_2}dt \left[ m{\dot{\vec r}}. {\dot{\vec{\bar r}}} +\frac{\gamma}{2}\left( {\dot{\vec r}}.{\ddot{\vec{\bar r}}}-{\ddot{\vec r}}.{\dot{\vec {\bar r}}}\right)-V(\vec r, \vec {\bar r})\right],
\end{equation}
where $\gamma=m\tau_0=2e^2/3c^3$.  It is then possible to identify
\begin{equation}
\label{09}
L= m{\dot{\vec r}}. {\dot{\vec{\bar r}}} +\frac{\gamma}{2}\left( {\dot{\vec r}}.{\ddot{\vec{\bar r}}}-{\ddot{\vec r}}.{\dot{\vec {\bar r}}}\right)-V(\vec r, \vec {\bar r})
\end{equation}
as the appropriate Lagrangian in the indirect representation. So, the system made of the radiation damping and of its time-reversed image globally behaves as a closed system. The Lagrangian (\ref{09}) can be written in a suggestive form by subtitution of the hyperbolic coordinates $\vec r_1$ and $\vec r_2$ \cite{Blasone} defined by 
\begin{equation}\label{10} \vec r = {1\over{\sqrt{2}}}\left(\vec r_{(1)} +\vec r_{(2)} \right); \;\; \vec {\bar r}
={1\over{\sqrt{2}}}\left(\vec r_{(1)} -\vec r_{(2)} \right)
\end{equation}
We find that the Lagrangian $L$ becomes
\begin{equation}\label{11} L={m\over 2}g_{ij}\; \dot{\vec r}_{(i)} .\dot{\vec r}_{(j)} -{\gamma
\over 2}\epsilon_{ij}\; \dot {\vec r}_{(i)} .\ddot{\vec r}_{(j)}-V[\vec r_{(1)}, \vec  r_{(2)}] \end{equation}
where the pseudo-euclidian metric $g_{ij}$ is given by
$g_{11}=-g_{22}=1$, $g_{12}=0$ and $\epsilon_{12}=-\epsilon_{21}
=1$. This Lagrangian is similar to the one discussed by Lukierski
et al \cite{Lukierski} (that is a special nonrelativistic limit of relativistic model of the particle with torsion investigated in \cite{MSP}), but in this case we have a pseudo-euclidian
metric. The  equations of motion corresponding to the Lagrangian
(\ref{11}) are
\begin{equation}\label{12} m\ddot {\vec r}_{(1)} - \gamma \stackrel{\ldots}{\vec r}_{(2)} =-\frac{\partial V}{\partial{\vec r}_{(2)}},\;\;
m\ddot {\vec r}_{(2)} -\gamma \stackrel{\ldots}{\vec r}_{(1)} =-\frac{\partial V}{\partial\vec r_{(1)}}. \end{equation}
On the hyperbolic plane, the equations (\ref{12}) shows that the dissipative term actully acts as a coupling between the systems $\vec r_{(1)}$ and $\vec r_{(2)}$. 

Recently, one of us, in \cite{Albert1} have studied the canonical quantization of the radiation damping. A Hamiltonian analysis is done in commutative and noncommutative scenarios, what leads to the quantization of the system, where the dynamical group structure associated with our system is that of $SU(1,1)$. In \cite{Albert2}, a supersymmetrized version of the model to the radiation damping, Eq.(\ref{11}), was developed. Its symmetries and the corresponding conserved Noether charges were discused. It is shown that this supersymmetric version provides a supersymmetric gene\-ralization of the Galilei algebra of the model \cite{Albert1}, where the supersymmetric action can be split into dynamically independent external and internal sectors.
 
\section{Concluding remarks}
\label{sec:3}

We have shown that in the pseudo-Euclidean metrics
the system made of a charge interacting with its own radiation and its time-reversed image, introduced by doubling the degrees of freedom as required by the
canonical formalism, actually behaves as a closed system described
by the Lagrangian (\ref{11}).
This formalism represents a new scenario in the study of this
very interesting system. The Lagrangian (\ref{11}) des\-cribes, in the hyperbolic plane, the dissipative
system of a charge interacting with its own radiation field, where the 2-labeled system represents the
reservoir or heat bath coupled to the 1-labed system.  Note that this Lagrangian is similar to the one discussed in \cite{Lukierski} (which is a
special nonrelativistic limit of relativistic model of the particle with torsion investigated in \cite{MSP}),
but in this case we have a pseudo-Euclidean metric and the radiation-damping constant, $\gamma$ , is
the coupling constant of a Chern-Simons-like term.
This formalism is important because it allows us to study the canonical quantization of the model (see Ref.\cite{Albert1}), and to study the symmetries of the model and their supersymmetric version (see Ref.\cite{Albert2}). In future works, we will study the introduction of gauge interactions into the model.

\section{Acknowledgement}
This work is supported  by CNPq Brazilian
Research Agency. In particular, ACRM  would like to acknowledge
the CNPq.

\end{document}